\documentclass{vgtc} 
\usepackage{amsmath, amsfonts, amsthm, mathtools, amssymb}

\usepackage{fancyhdr}
\usepackage{standalone}

\graphicspath{{Images/}{./}} 

\usepackage{times}                     

\usepackage{mathptmx}                  

\input{IsoComponentFiltering.sty}

\onlineid{0}

\vgtccategory{Research}

\title{Filtering of Small Components for Isosurface Generation}

\author{Devin Zhao\thanks{e-mail: zhao.3657@buckeyemail.osu.edu}\\ %
        \scriptsize The Ohio State University %
\and Rephael Wenger\thanks{e-mail: wenger.4@osu.edu}\\ %
     \parbox{1.4in}{\scriptsize \centering The Ohio State University}}

\teaser{
  \centering
  \begin{subfigure}{0.2\textwidth}
  \includegraphics[height=100px]{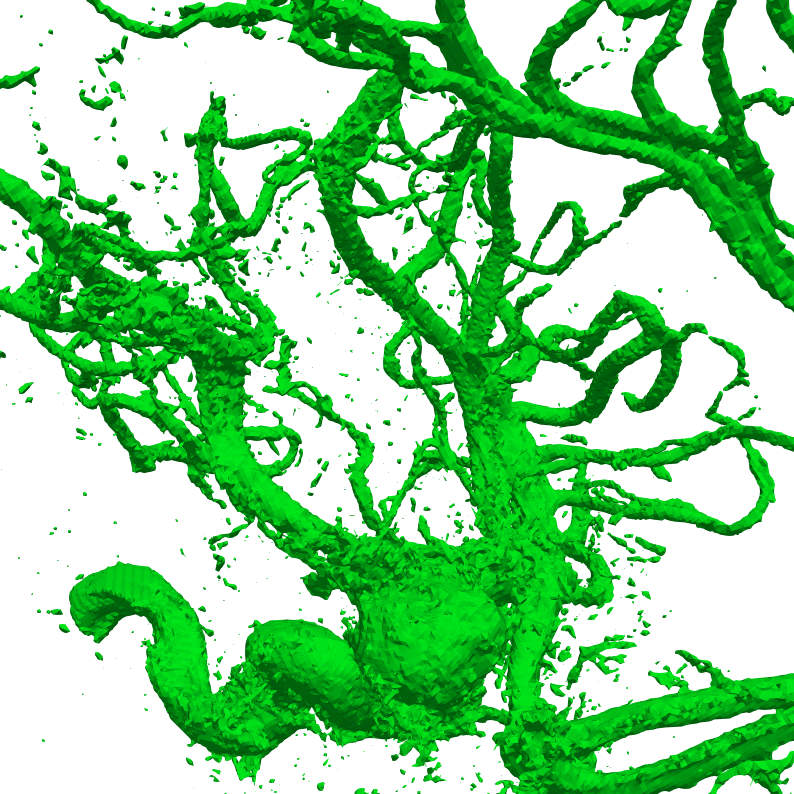}
  \caption{}
  \end{subfigure}
  \quad
  \begin{subfigure}{0.2\textwidth}
  \includegraphics[height=100px]{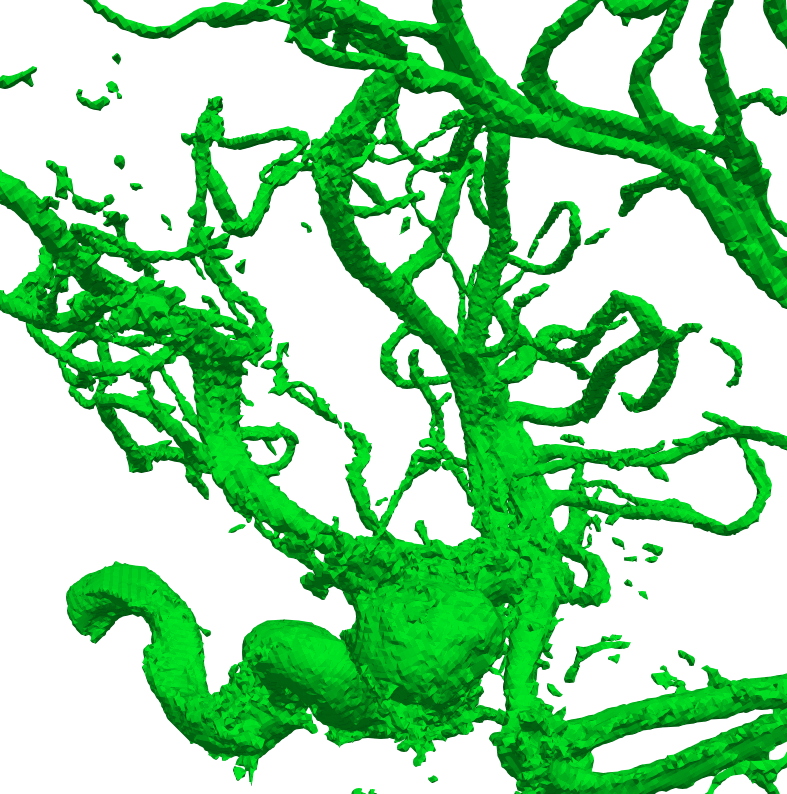}
  \caption{}
  \end{subfigure}
  \quad
  \begin{subfigure}{0.2\textwidth}
  \includegraphics[height=100px]{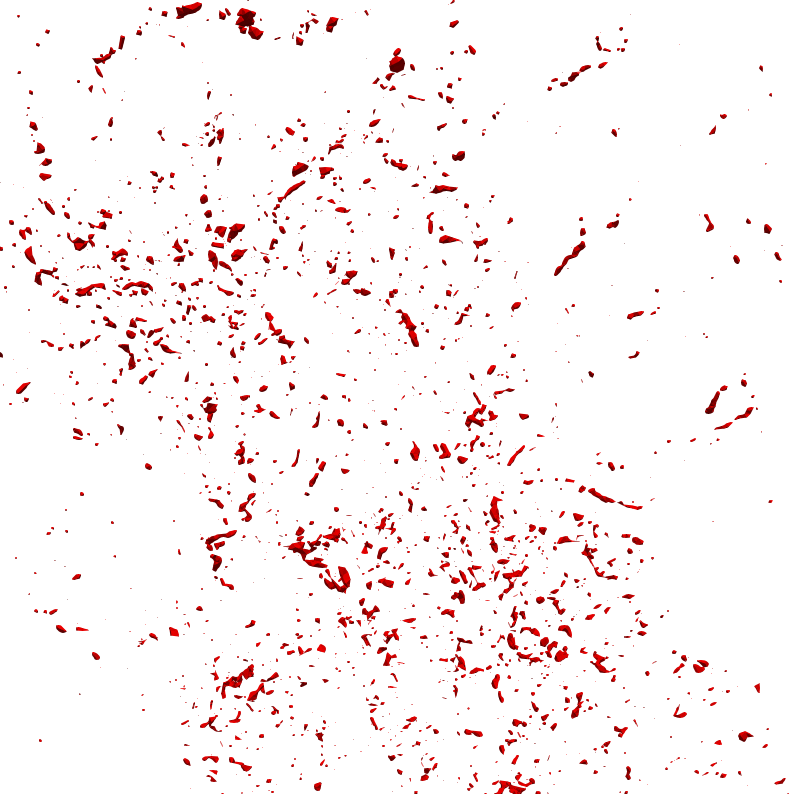}
  \caption{}
  \end{subfigure}
  \quad
  \begin{subfigure}{0.2\textwidth}
  \includegraphics[height=100px]{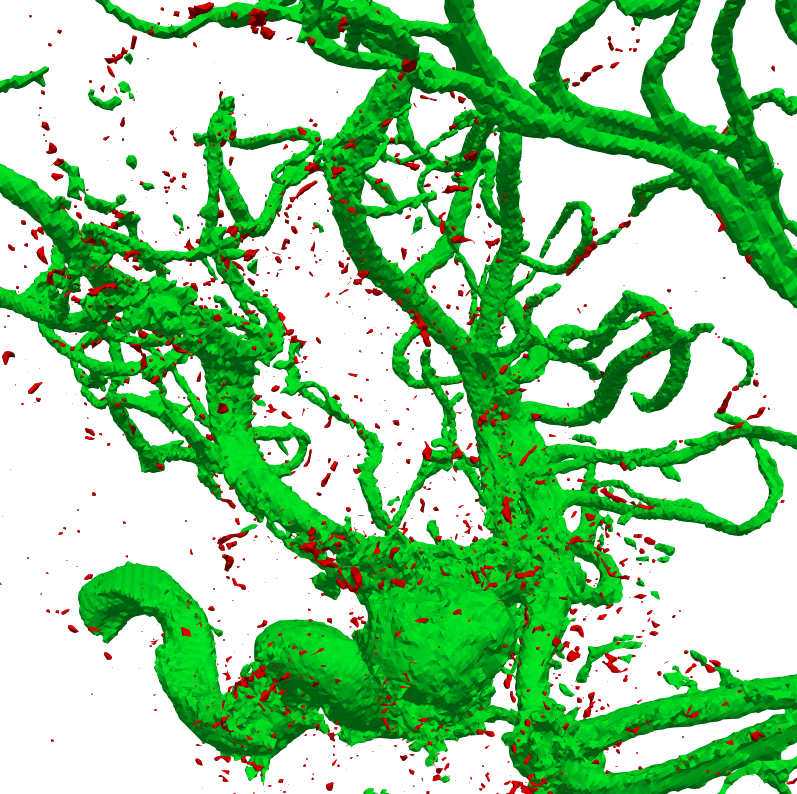}
  \caption{}
  \end{subfigure}
  \caption{All four images are produced from a CT scan
           of a brain aneurysm~\cite{open-scivis-datasets}
           and generated with isovalue 30.5. 
  (a) Original isosurface without filtering.
  (b) Isosurface after filtering out components of size less than or equal to 5.
  The filtered isosurface has 322 connected components.
  (c) All of the 3921 connected components that were removed from the original isosurface.
  (d) Images (b) and (c) combined to recreate the original isosurface.
  }
  \label{fig:teaser}
}

\abstract{
Let $f: \Rthree \rightarrow \RR$ be a scalar field.
An isosurface is a piecewise linear approximation
of a level set $f^{-1}(\sigma)$ for some $\sigma \in \RR$
built from some regular grid sampling of $f$.
Isosurfaces constructed from scanned data such as CT scans or MRIs
often contain extremely small components
that distract from the visualization
and do not form part of any geometric model produced from the data.
Simple prefiltering of the data can remove such small components
while having no effect on the large components
that form the body of the visualization.
We present experimental results on such filtering.
} 


\begin{document}


\firstsection{Introduction}
\labelsec{intro}

\maketitle

\thispagestyle{fancy}

Given a regular grid sampling 
of a scalar field $f: \mathbb{R}^3 \rightarrow \mathbb{R}$,
an isosurface is a piecewise linear approximation
of a level set $f^{-1}(\sigma)$ for some $\sigma \in \mathbb{R}$.
Isosurfaces are commonly used to visualize region boundaries
in 3D datasets and as a step in building geometric
models from those datasets.
The Marching Cubes algorithm 
and its numerous variants~\cite{lorensen1987marchingcubes,
wenger2013isosurfaces} quickly construct
an isosurface from a regular grid sampling of a scalar field.

The Marching Cubes algorithm constructs an isosurface
that is extremely faithful to small details
of the level set $f^{-1}(\sigma)$
up to the resolution of the regular grid sampling.
This faithfulness has the advantage of not masking or filtering small features
that are present in the sample data.
However, the faithfulness to detail has the corresponding disadvantage
of representing noise that is present in the sample data.
In particular, isosurfaces constructed from scanned data
such as CT scans or MRIs often contain such noise.
Convolution filters can be used to suppress or remove such noise,
but at the cost of modifying all the scalar data,
reducing the fidelity of the isosurface to the original data.

Applying Marching Cubes to noisy scalar data often results
in small connected components in the isosurface.
Such small connected components distract from the visualization.
They also do not contribute to any geometric models
built from the isosurface.

Removing small components from a geometric model
requires building some type of connectivity representation
between the mesh elements of that model.
While constructing such a representation is certainly possible,
a simpler approach is to identify small connected components 
in the scalar grid,
modifying only those parts of the scalar grid that generate small components.

More precisely, let $\superGsig$ be the subgraph
of the regular grid induced 
by the set of grid vertices $\{v: f(v) \ge \sigma\}$.
Subgraph $\superGsig$ represents the set $\{x : f(x) \ge \sigma\}$,
the \defterm{superlevel} set of $f$ for value $\sigma$.
We identify ``small'' connected components $\mu_i$ of $\superGsig$,
and ``remove'' them
by setting the scalar values of $v \in \mu_i$  to be below $\sigma$.
By changing the scalar values of $\{v: v \in \mu_i\}$ to be below $\sigma$,
we eliminate the small isosurface components that surround the small $\mu_i$.
Similarly, we identify ``small''
connected components of the subgraph $\subGsig$ 
induced by $\{v : f(v) \le \sigma\}$
and change their scalar values to be above $\sigma$.

Following numerous image processing papers on region growing and segmentation,
we use a union-find data structure to quickly construct $\subGsig$
and $\superGsig$ and identify their small components.

In this paper, we present the following:
\begin{enumerate}
\item A simple algorithm to remove small isosurface components 
by identifying and removing small components
of the subgraphs induced by $\{v: f(v) \le \sigma\}$
and $\{v: f(v) \ge \sigma\}$.
\item Extensive experimental results on applying the algorithm
to scalar data sets.
\end{enumerate}

While the algorithm we present is similar 
to many region growing and image segmentation algorithms,
our contribution lies in reporting the results
of applying this algorithm to isosurface construction
of numerous volumetric data sets.

\SetKwFunction{bwareaopen}{bwareaopen}
\SetKwFunction{connectedComponentsWithStats}{connectedComponentsWithStats}
\SetKwFunction{removeSmallObjects}{remove\_small\_objects}
\SetKwFunction{connectedComponentsIIID}{connected-components-3D}
\SetKwFunction{vtkConnectivityFilter}{vtkConnectivityFilter}
\SetKwFunction{connectedComponentImageFilter}{connectedComponentImageFilter}

\section{Previous Work}
\labelsec{previous}

Numerous papers discuss filtering and segmenting images and volumes 
based on connected components.
Seeded region growing algorithms grow regions 
(connected components) based on the similarity between pixels~\cite{adams1994seeded,fan2005seeded,mehnert1997improved,wan2003symmetric}.
Watershed algorithms segment images
by splitting images into connected components of sublevel sets~\cite{bieniek2000efficient,kornilov2018overview,vincent1991watersheds}.

Filtering based on connected components of scalar value is discussed
in~\cite{jones1999connected,serra2002connection,serra2006lattice,xu2015connected}.
Connected component based filtering has been applied 
to material microstructures~\cite{patel2022topfilter},
concrete crack detection~\cite{cao2023continuous},
detection of astronomical objects~\cite{teeninga2016statistical},
electrophoresis gels~\cite{kazhiyur2006contour},
medical CT scans~\cite{lisnichenko20223d}
and medical MRI images~\cite{reddy2023brain}.

Fast algorithms for forming connected components generally
rely upon union-find data structures.
(See~\cite{cormen2022alg} for descriptions and implementations
of union-find data structures.)
Applying union-find to 2D and 3D segmentation based
on region growing is described in~\cite{fiorio1996union,fiorio2000union}.

Most visualization packages have some routine for computing
connected components of images and 3D regular grids.
The Matlab~\cite{matlab} $\bwareaopen$ function 
and the Python SciKit~\cite{scikitImage} $\removeSmallObjects$
remove small components from binary (0 and 1) 2D images and 3D volumes.
The OpenCV~\cite{opencv} $\connectedComponentsWithStats$ function computes connected components
of 2D images as well as statistics such as area, centroid, and bounding rectangles
of those components.
Python's SciKit~\cite{scikitImage} $\connectedComponentsIIID$ function 
provides similar functionality for 3D volumes.
The Visualization Toolkit~\cite{vtk} $\vtkConnectivityFilter$
identifies the connected component containing some seed voxel,
where connectivity can be based on scalar values.
The Insight Toolkit (ITK)~\cite{itk} $\connectedComponentImageFilter$
computes connected components in binary (0 and 1) 2D images and volumes.

\section{Small Component Filtering}
\labelsec{algorithm}

\begin{algorithm}[t]
\caption{Algorithm merging vertices into the same disjoint set if they are in the same component}
\labelalg{merge_components}
\NoLineNum{\Procedure \MergeComponents{$DS$,$F$, $\sigma$}}
\tcc{$DS$ is the set of grid vertices}
\tcc{$F$ is a 3D array of scalar values}
\tcc{$\sigma$ is an isovalue}
\ForEach{$v \in DS$}{
\lIf{($F[v] \le \sigma$)}{
\tcc{Create set $\{v\}$}
\MakeSet{$v$}
}
}
\tcc{Iterate over $x$, $y$, $z$ directions}
\For{$d = 0,1,2$}{
\ForEach{grid edge $(v,v')$ with direction $d$}{
\If{(($F[v] \le \sigma$) \KwAnd ($F[v'] \le \sigma$)) \KwOr (($F[v] \ge \sigma$) \KwAnd ($F[v'] \ge \sigma$)) }{
\tcc{Union the sets containing $v$ and $v'$}
\tcc{Compute and store the set size at the ``root'' vertex}
\Union{v,v'}\;
}
}
}
\end{algorithm}

Consider $\superGsig$, the subgraph of the grid containing 
all lattice points $\{ v: f(v) \geq \sigma\}$.
We use 6-connectivity, where each grid vertex is connected
to the grid vertices directly above/below, left/right, before/after it. 

As in \cite{fiorio1996union,fiorio2000union,patel2022topfilter}
(and many other papers,)
we use the union-find data structure to identify the connected components
of $\superGsig$.
We start by forming the set $\{v\}$ for each $v \in \superGsig$.
For every grid edge $\{v,v'\}$ where $v,v' \in \superGsig$,
we form the union of the sets containing $v$ and $v'$.
With each set,
we also keep count of the number of elements in the set.

\bigskip
\noindent 
For each connected component $\mu_i$ of $\superGsig$
whose size (number of vertices) is less than a threshold,
we assign new scalar values to the vertices $v \in \mu_i$
as follows:
\begin{itemize}
\item Let $(x_v,y_v,z_v)$ be the coordinates of grid vertex $v$.
\item Let $v_x$ and $v'_x$ be the closest vertices
to the left and right of $v$ in the row $(*,y_v,z_v)$
whose scalar values are below $\sigma$.
\item Similarly, let $v_y$ and $v'_y$ be the closest
vertices below and above $v$ in the column $(x_v,*,z_v)$
whose scalar values are below $\sigma$
and let $v_z$ and $v'_z$ be the closest
vertices before and after $v$ in $(x_v,y_v,*)$
whose scalar values are below $\sigma$.
\item Set the scalar value of $v$ to be the average
of the scalar values of $\{v_x,v'_x,v_y,v'_y,v_z,v'_z\}$.
\end{itemize}

Consider $\subGsig=\{v: f(v) \leq \sigma\}$ and $\Gamma_{\sigma}^{+-}=\{v: f(v) \leq \sigma \text{ or } f(v) \geq \sigma\}$. We can apply a similar algorithm described above to modify the scalar values of each vertex in every $\mu_i$.

\begin{figure}
\includegraphics[height=125px]{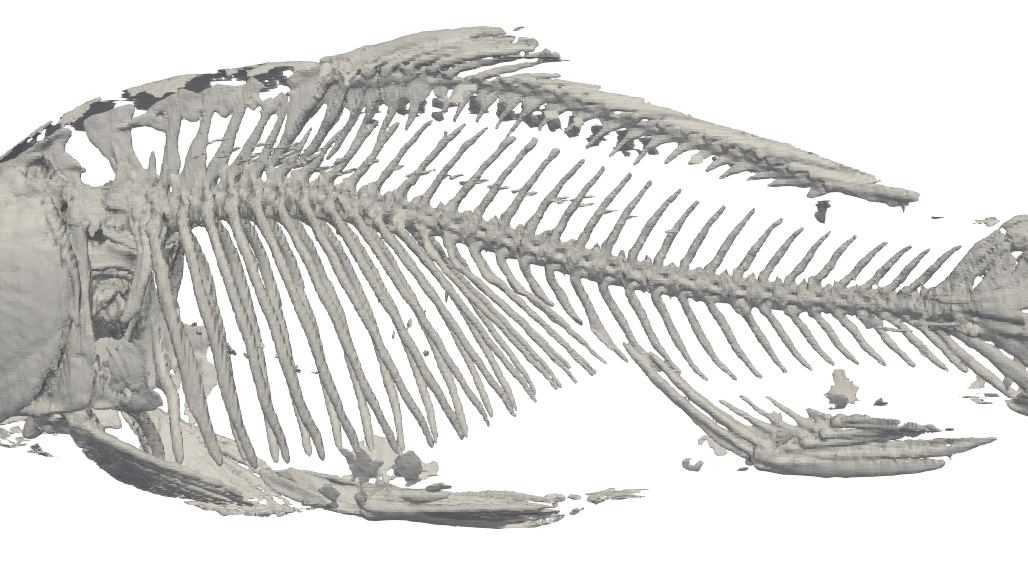}
\includegraphics[height=120px]{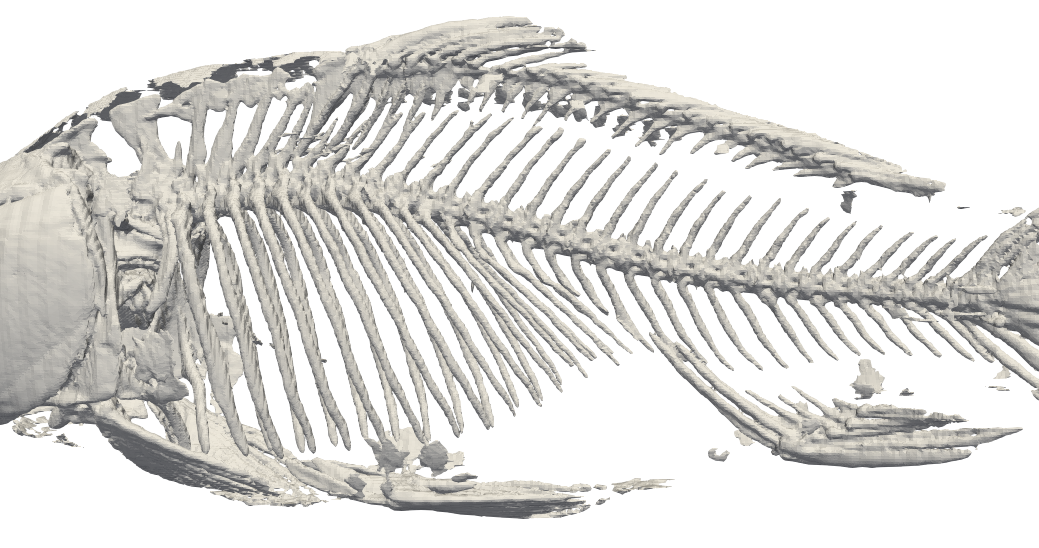}
\includegraphics[height=120px]{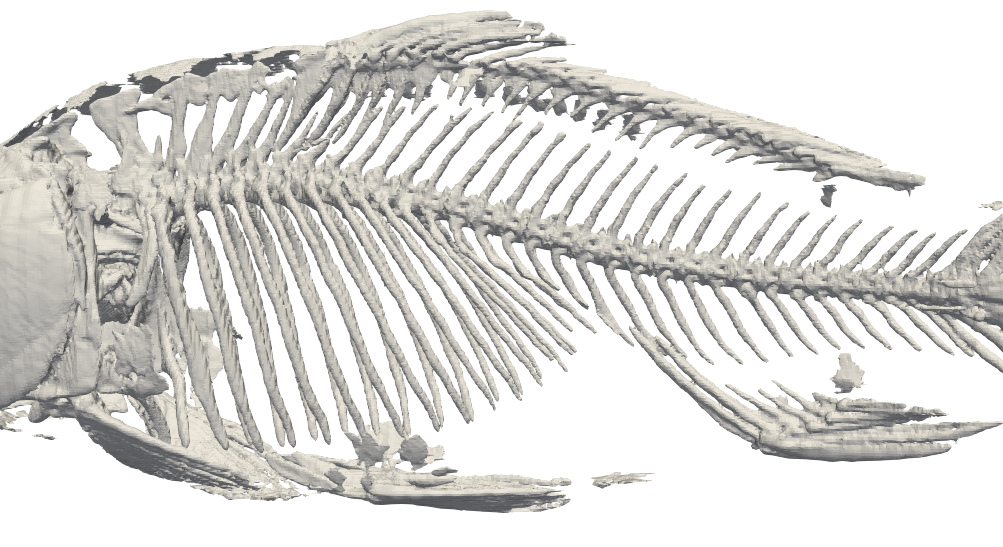}
\caption{Isosurfaces of a carp skeleton after filtering out all components of size 10, 20 and 50 respectively. Generated with isovalue 1150.5.}
\end{figure}

\section{Experimental Results}
\labelsec{experimental}

\begin{figure}
\includegraphics[height=130px]{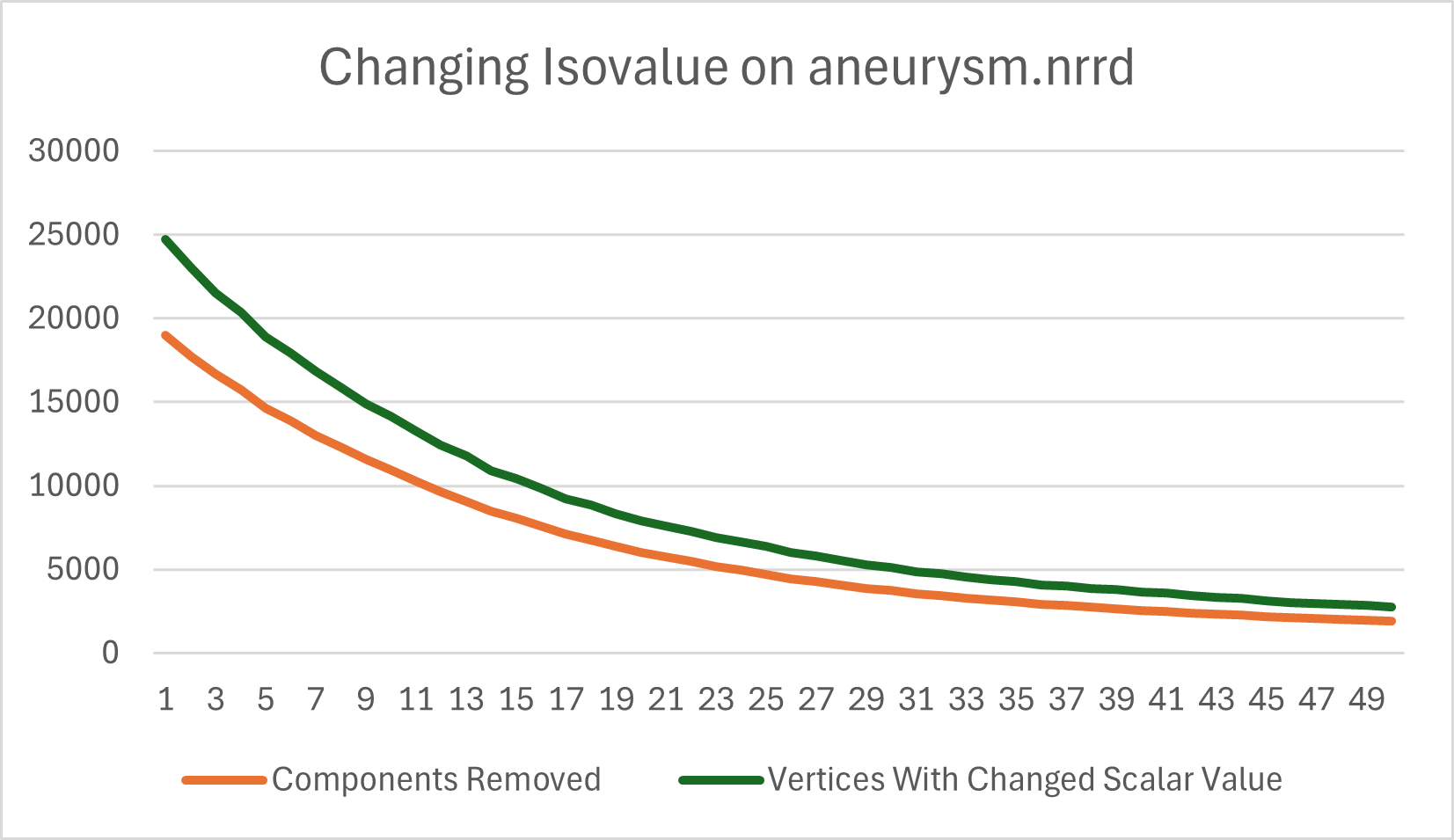}
\includegraphics[height=150px]{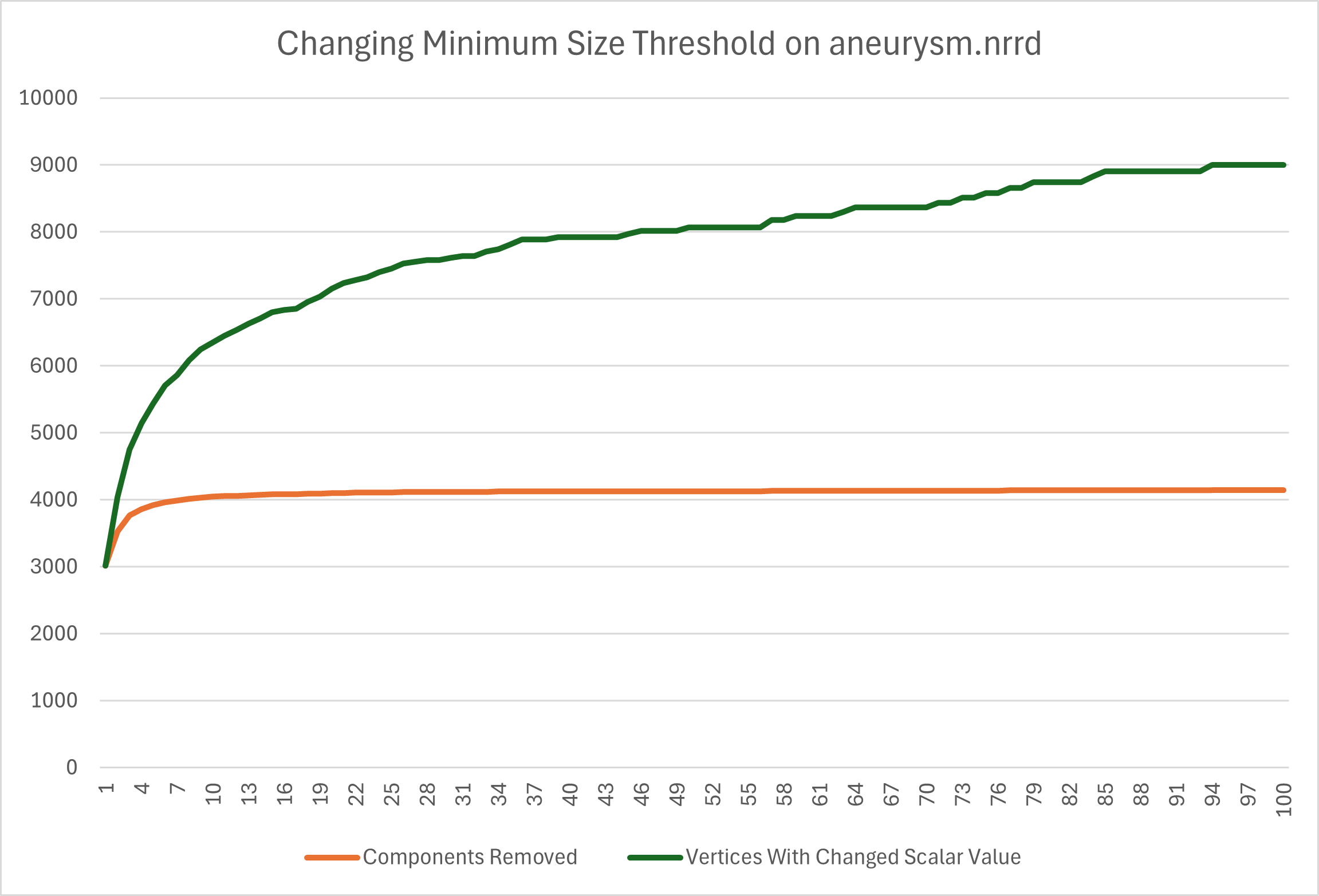}
\caption{Top: Comparing different isovalues ($x$-axis) to the number of components removed and scalar values changed (y-axis) on aneurysm. Done with a minimum size threshold of 5. Bottom: Comparing different minimum size thresholds (x-axis) to the number of components removed and scalar values changed (y-axis). Done with an isovalue of 30.5.}
\label{fig:aneurysm}
\end{figure}

\begin{figure}
\includegraphics[height=150px]{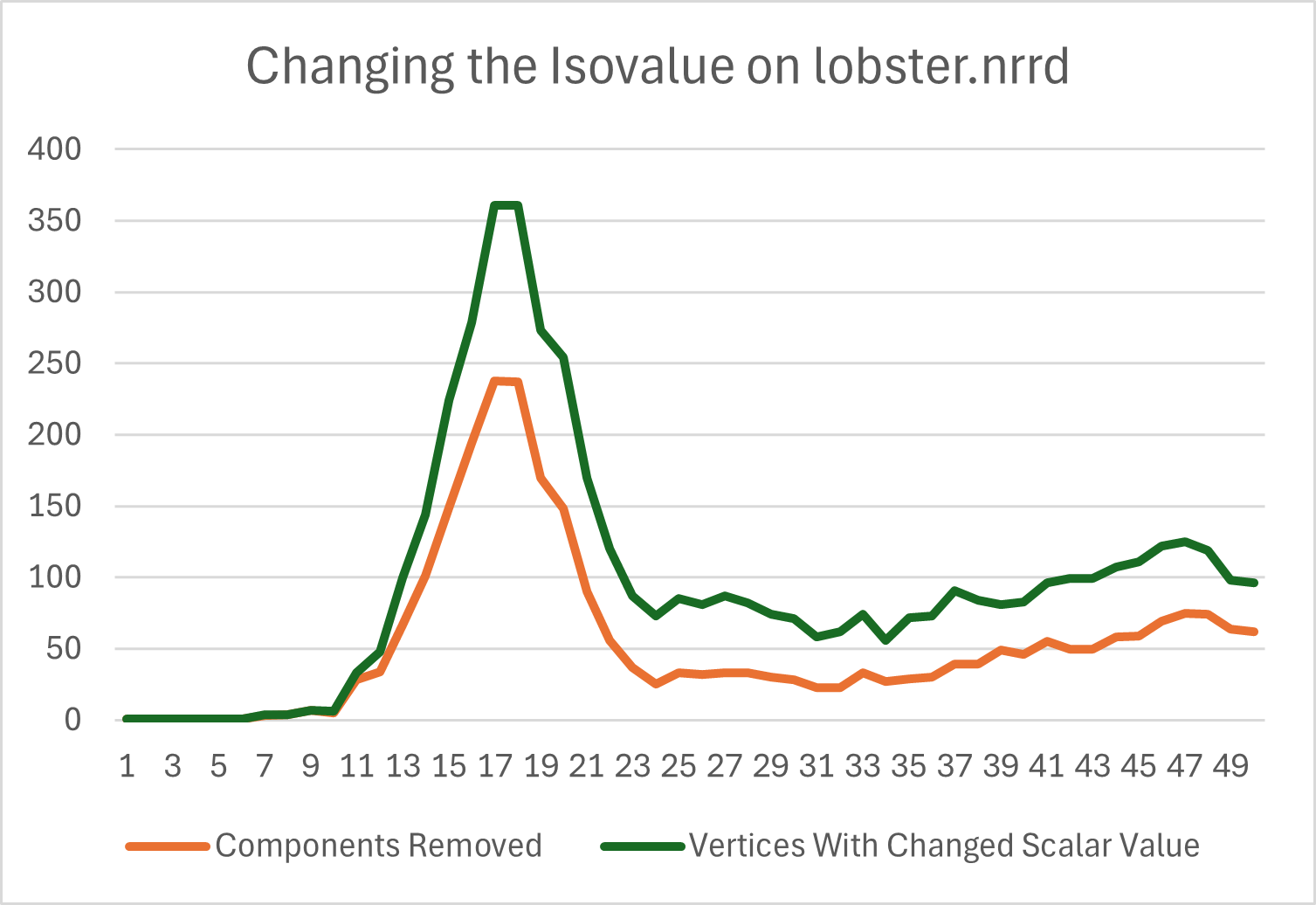}
\includegraphics[height=150px]{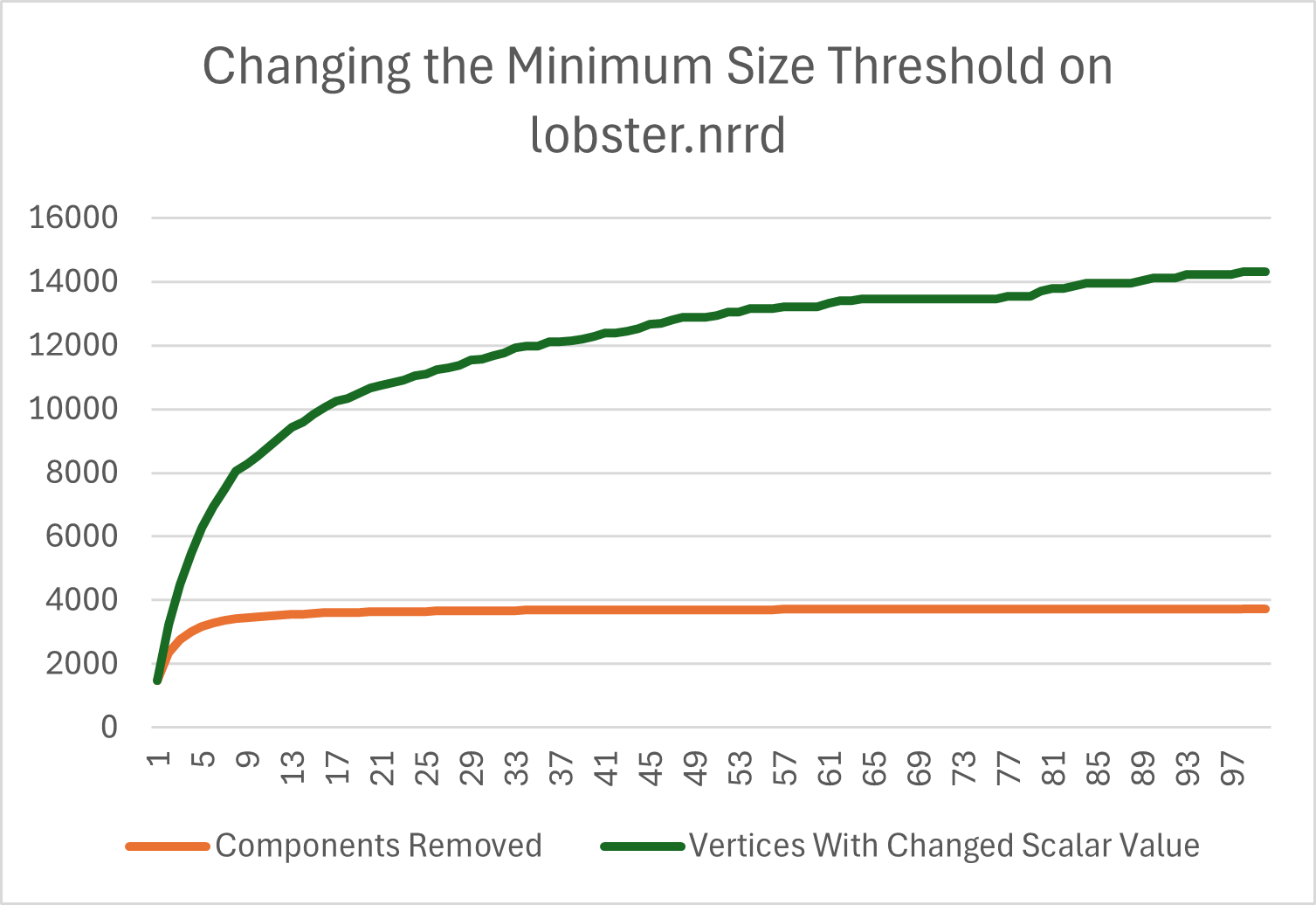}
\caption{Top: Comparing different isovalues to the number of components removed and scalar values changed on lobster. Done with a minimum size threshold of 5. Bottom: Comparing different minimum size thresholds to the number of components removed and scalar values changed. Done with an isovalue of 20.5.}
\label{fig:lobster}
\end{figure}

\begin{figure}
\includegraphics[height=130px]{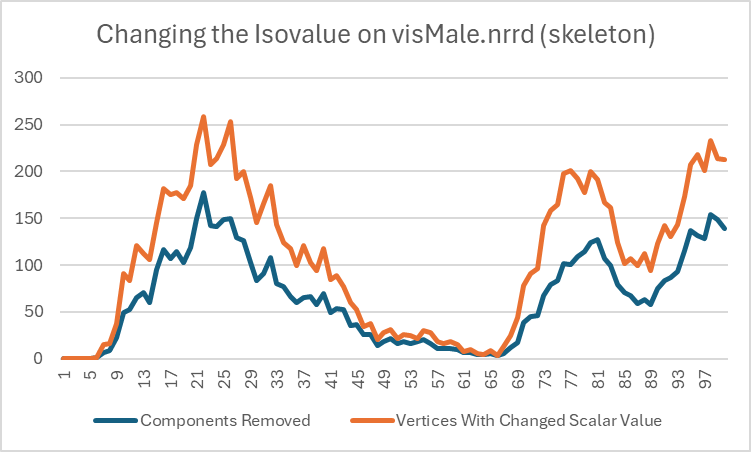}
\includegraphics[height=130px]{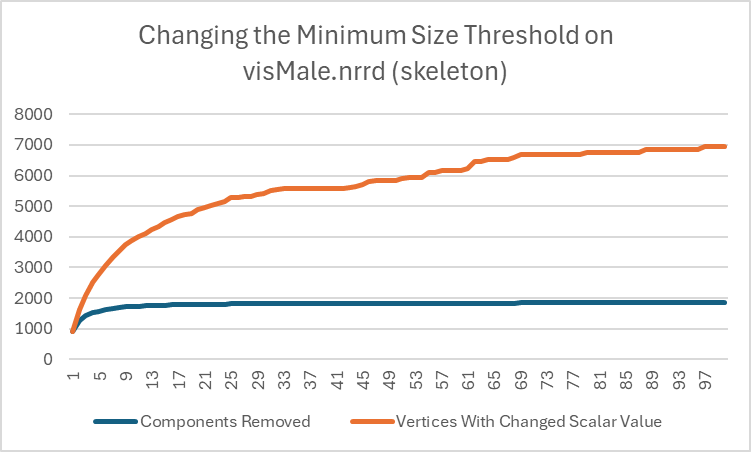}
\caption{Top: Comparing different isovalues to the number of components removed and scalar values changed on visMale. Done with a minimum size threshold of 5. Bottom: Comparing different minimum size thresholds to the number of components removed and scalar values changed. Done with an isovalue of 70.5.}
\label{fig:vismale}
\end{figure}

We experimented on volumetric datasets from~\cite{open-scivis-datasets}.
Figures~\ref{fig:aneurysm}, \ref{fig:lobster}, and~\ref{fig:vismale}
compare the number of components removed and vertex scalar values modified 
as the minimum size threshold changes
or as the isovalue changes.
Figures~\ref{fig:aneurysm}, \ref{fig:lobster}, and~\ref{fig:vismale}
analyze isosurfaces from the datasets aneurymn, lobster, or vismale,
respectively.

One common trend on the Minimum Size versus Components Removed graphs (Figures \ref{fig:aneurysm}, \ref{fig:lobster}, \ref{fig:vismale}) is that the curve starts flattening out at around a filter size of 20. Tables~\ref{table:filterSize} and \ref{table:filterSizeII} also suggest a similar trend. The difference in the percentage of components removed using filter sizes of 10 versus 20 is slightly larger than the difference between filter sizes of 20 versus 50 (which is at most 5\% for all datasets) for the majority of the datasets, showing that the amount of components filtered out starts slowing down after 20. This indicates that 20 would be a reasonable number to set the minimum filter size to, and deleting any components larger than that would lead to removing meaningful parts of the isosurface. 

Additionally, some of the Isovalue vs Components Removed figures above contain spikes, showing that certain isovalues produce a lot of noise in the resulting isosurface. While this does not say much about the effectiveness of our filtering algorithm, we can use this graph to estimate an appropriate isovalue to generate our isosurface with. If we know that the object is relatively clean and only consists of a few large components, then we can pick an isovalue that is at the bottom of the graph.

Tables~\ref{table:filterSize} and~\ref{table:filterSizeII}
present results of removing small components
of scalar values that are GREATER than or equal to the isovalue.
A component was removed only if the component size was the
less than or equal to the given threshold (filter size).
Table~\ref{table:filterSize} presents the number of components removed
and the number of scalar values modified
for filter sizes 1, 5, and 10.
Table~\ref{table:filterSizeII} presents the same measurements
for filter sizes 20, and 50.

Tables~\ref{table:filterSizeIII} and~\ref{table:filterSizeIV}
present results of removing small components
of scalar values that are LESS than or equal to the isovalue.
Table~\ref{table:filterSizeIII} presents results for filter sizes
1, 5, and 10,
and Table~\ref{table:filterSizeIV} presents results 
for filter sizes 20, and 50.

\begin{table*}
\centering
\begin{threeparttable}
\centering
\begin{minipage}{\textwidth}
\begin {tabular}{lrrrrrrrrrr}%
\toprule & & \# Active & Filter & Total \# & \multicolumn {2}{c}{Filter size 1} & \multicolumn {2}{c}{Filter size 5} & \multicolumn {2}{c}{Filter size 10} \\ Dataset & Isovalue & cubes & type & comp. & Comp. & Scalar & Comp. & Scalar & Comp. & Scalar \\ & & & & & removed & values & removed & values & removed & values \\ & & & & & & modified & & modified & & modified \\ \toprule %
\midrule abdominal\_stent&1350.5&488K&$\geq \sigma $&2649&1317 (49.7\%)&1317&2122 (80.1\%)&3658&2302 (86.9\%)&4987\\%
\midrule aneurysm&30.5&163K&$\geq \sigma $&4241&3015 (71.1\%)&3015&3921 (92.5\%)&5434&4034 (95.1\%)&6243\\%
\midrule bonsai&50.5&305K&$\geq \sigma $&1428&338 (23.7\%)&338&780 (54.6\%)&1639&928 (65.0\%)&2752\\%
\midrule carp&600.5\tnote {\fnII }&450K&$\geq \sigma $&81&4 (4.9\%)&4&5 (6.2\%)&6&5 (6.2\%)&6\\%
\midrule carp&1150.5\tnote {\fnIII }&662K&$\geq \sigma $&2358&948 (40.2\%)&948&1721 (73.0\%)&3112&1917 (81.3\%)&4595\\%
\midrule colon\_prone&1500.5&1433K&$\geq \sigma $&7286&3867 (53.1\%)&3867&6061 (83.2\%)&9931&6463 (88.7\%)&12951\\%
\midrule colon\_supine&1500.5&1392K&$\geq \sigma $&6734&3470 (51.5\%)&3470&5513 (81.9\%)&9135&5896 (87.6\%)&12009\\%
\midrule lobster&20.5&239K&$\geq \sigma $&4031&1462 (36.3\%)&1462&3171 (78.7\%)&6262&3482 (86.4\%)&8520\\%
\midrule MRIwoman&1100.5&599K&$\geq \sigma $&8788&4921 (56.0\%)&4921&6448 (73.4\%)&8936&6603 (75.1\%)&10099\\%
\midrule skull&40.5&959K&$\geq \sigma $&4819&2965 (61.5\%)&2965&4055 (84.1\%)&5939&4232 (87.8\%)&7262\\%
\midrule visMale&55.5\tnote {\fnII }&279K&$\geq \sigma $&283&26 (9.2\%)&26&44 (15.5\%)&80&53 (18.7\%)&152\\%
\midrule visMale&70.5\tnote {\fnIII }&453K&$\geq \sigma $&2057&904 (43.9\%)&904&1577 (76.7\%)&2775&1724 (83.8\%)&3869\\\bottomrule %
\end {tabular}%

  \begin{tablenotes}
\item [\fnII] Isovalue for skin.
\item [\fnIII] Isovalue for skeleton.
\end{tablenotes}
\end{minipage}
\caption{Number of components removed and scalar values modified across various datasets using different minimum size thresholds. All removed components contain vertices with scalar value greater than or equal to the isovalue.}
\label{table:filterSize}
\end{threeparttable}
\end{table*}

\begin{table*}
\centering
\begin{threeparttable}
\centering
\begin{minipage}{\textwidth}
  \begin {tabular}{lrrrrrrrr}%
\toprule & & \# Active & Filter & Total \# & \multicolumn {2}{c}{Filter size 20} & \multicolumn {2}{c}{Filter size 50} \\ Dataset & Isovalue & cubes & type & comp. & Comp. & Scalar & Comp. & Scalar\\ & & & & & removed & values & removed & values\\ & & & & & & modified & & modified \\ \toprule %
\midrule abdominal\_stent&1350.5&488K&$\geq \sigma $&2649&2437 (92.0\%)&6977&2518 (95.1\%)&9490\\%
\midrule aneurysm&30.5&163K&$\geq \sigma $&4241&4099 (96.7\%)&7154&4130 (97.4\%)&8066\\%
\midrule bonsai&50.5&305K&$\geq \sigma $&1428&1057 (74.0\%)&4614&1141 (79.9\%)&7361\\%
\midrule carp&600.5\tnote {\fnII }&450K&$\geq \sigma $&81&5 (6.2\%)&6&6 (7.4\%)&29\\%
\midrule carp&1150.5\tnote {\fnIII }&662K&$\geq \sigma $&2358&2016 (85.5\%)&6034&2090 (88.6\%)&8327\\%
\midrule colon\_prone&1500.5&1433K&$\geq \sigma $&7286&6660 (91.4\%)&15791&6813 (93.5\%)&20426\\%
\midrule colon\_supine&1500.5&1392K&$\geq \sigma $&6734&6096 (90.5\%)&14904&6212 (92.2\%)&18648\\%
\midrule lobster&20.5&239K&$\geq \sigma $&4031&3633 (90.1\%)&10669&3703 (91.9\%)&12891\\%
\midrule MRIwoman&1100.5&599K&$\geq \sigma $&8788&6680 (76.0\%)&11236&6713 (76.4\%)&12363\\%
\midrule skull&40.5&959K&$\geq \sigma $&4819&4319 (89.6\%)&8552&4396 (91.2\%)&10928\\%
\midrule visMale&55.5\tnote {\fnII }&279K&$\geq \sigma $&283&55 (19.4\%)&181&61 (21.6\%)&371\\%
\midrule visMale&70.5\tnote {\fnIII }&453K&$\geq \sigma $&2057&1795 (87.3\%)&4896&1827 (88.8\%)&5837\\\bottomrule %
\end {tabular}%

  \begin{tablenotes}
\item [\fnII] Isovalue for skin.
\item [\fnIII] Isovalue for skeleton.
\end{tablenotes}
\end{minipage}
\caption{(Table 1 continued) Number of components removed and scalar values modified across various datasets using different minimum size thresholds. All removed components contain vertices with scalar value greater than or equal to the isovalue.}
\label{table:filterSizeII}
\end{threeparttable}
\end{table*}

\begin{table*}
\centering
\begin{threeparttable}
\centering
\begin{minipage}{\textwidth}
  \begin {tabular}{lrrrrrrrrrr}%
\toprule & & \# Active & Filter & Total \# & \multicolumn {2}{c}{Filter size 1} & \multicolumn {2}{c}{Filter size 5} & \multicolumn {2}{c}{Filter size 10} \\ Dataset & Isovalue & cubes & type & comp. & Comp. & Scalar & Comp. & Scalar & Comp. & Scalar \\ & & & & & removed & values & removed & values & removed & values \\ & & & & & & modified & & modified & & modified \\ \toprule %
\midrule abdominal\_stent&1350.5&488K&$\leq \sigma $&2649&42 (1.6\%)&42&62 (2.3\%)&94&65 (2.5\%)&122\\%
\midrule aneurysm&30.5&163K&$\leq \sigma $&4241&67 (1.6\%)&67&85 (2.0\%)&121&85 (2.0\%)&121\\%
\midrule bonsai&50.5&305K&$\leq \sigma $&1428&61 (4.3\%)&61&125 (8.8\%)&261&153 (10.7\%)&483\\%
\midrule carp&600.5\tnote {\fnII }&450K&$\leq \sigma $&81&9 (11.1\%)&9&27 (33.3\%)&62&33 (40.7\%)&113\\%
\midrule carp&1150.5\tnote {\fnIII }&662K&$\leq \sigma $&2358&71 (3.0\%)&71&141 (6.0\%)&273&163 (6.9\%)&448\\%
\midrule colon\_prone&1500.5&1433K&$\leq \sigma $&7286&183 (2.5\%)&183&262 (3.6\%)&384&276 (3.8\%)&486\\%
\midrule colon\_supine&1500.5&1392K&$\leq \sigma $&6734&188 (2.8\%)&188&282 (4.2\%)&457&296 (4.4\%)&568\\%
\midrule lobster&20.5&239K&$\leq \sigma $&4031&132 (3.3\%)&132&229 (5.7\%)&415&246 (6.1\%)&544\\%
\midrule MRIwoman&1100.5&599K&$\leq \sigma $&8788&1584 (18.0\%)&1584&2000 (22.8\%)&2646&2026 (23.1\%)&2839\\%
\midrule skull&40.5&959K&$\leq \sigma $&4819&252 (5.2\%)&252&317 (6.6\%)&419&326 (6.8\%)&480\\%
\midrule visMale&55.5\tnote {\fnII }&279K&$\leq \sigma $&283&67 (23.7\%)&67&124 (43.8\%)&224&145 (51.2\%)&390\\%
\midrule visMale&70.5\tnote {\fnIII }&453K&$\leq \sigma $&2057&88 (4.3\%)&88&161 (7.8\%)&301&177 (8.6\%)&422\\\bottomrule %
\end {tabular}%

  \begin{tablenotes}
\item [\fnII] Isovalue for skin.
\item [\fnIII] Isovalue for skeleton.
\end{tablenotes}
\end{minipage}
\caption{Number of components removed and scalar values modified across various datasets using different minimum size thresholds. All removed components contained vertices with scalar value less than or equal to the isovalue. All removed components contain vertices with scalar value less than or equal to the isovalue.}
\label{table:filterSizeIII}
\end{threeparttable}
\end{table*}

\begin{table*}
\centering
\begin{threeparttable}
\centering
\begin{minipage}{\textwidth}
  \begin {tabular}{lrrrrrrrr}%
\toprule & & \# Active & Filter & Total \# & \multicolumn {2}{c}{Filter size 20} & \multicolumn {2}{c}{Filter size 50} \\ Dataset & Isovalue & cubes & type & comp. & Comp. & Scalar & Comp. & Scalar\\ & & & & & removed & values & removed & values\\ & & & & & & modified & & modified \\ \toprule %
\midrule abdominal\_stent&1350.5&488K&$\leq \sigma $&2649&66 (2.5\%)&142&66 (2.5\%)&142\\%
\midrule aneurysm&30.5&163K&$\leq \sigma $&4241&85 (2.0\%)&121&85 (2.0\%)&121\\%
\midrule bonsai&50.5&305K&$\leq \sigma $&1428&180 (12.6\%)&869&197 (13.8\%)&1420\\%
\midrule carp&600.5\tnote {\fnII }&450K&$\leq \sigma $&81&41 (50.6\%)&235&55 (67.9\%)&725\\%
\midrule carp&1150.5\tnote {\fnIII }&662K&$\leq \sigma $&2358&171 (7.3\%)&568&188 (8.0\%)&1146\\%
\midrule colon\_prone&1500.5&1433K&$\leq \sigma $&7286&278 (3.8\%)&510&278 (3.8\%)&510\\%
\midrule colon\_supine&1500.5&1392K&$\leq \sigma $&6734&301 (4.5\%)&641&302 (4.5\%)&665\\%
\midrule lobster&20.5&239K&$\leq \sigma $&4031&264 (6.5\%)&801&276 (6.8\%)&1141\\%
\midrule MRIwoman&1100.5&599K&$\leq \sigma $&8788&2039 (23.2\%)&3031&2040 (23.2\%)&3053\\%
\midrule skull&40.5&959K&$\leq \sigma $&4819&330 (6.8\%)&537&331 (6.9\%)&561\\%
\midrule visMale&55.5\tnote {\fnII }&279K&$\leq \sigma $&283&161 (56.9\%)&643&181 (64.0\%)&1280\\%
\midrule visMale&70.5\tnote {\fnIII }&453K&$\leq \sigma $&2057&186 (9.0\%)&551&191 (9.3\%)&703\\\bottomrule %
\end {tabular}%

  \begin{tablenotes}
\item [\fnII] Isovalue for skin.
\item [\fnIII] Isovalue for skeleton.
\end{tablenotes}
\end{minipage}
\caption{(Table 3 continued) Number of components removed and scalar values modified across various datasets using different minimum size thresholds. All removed components contain vertices with scalar value less than or equal to the isovalue.}
\label{table:filterSizeIV}
\end{threeparttable}
\end{table*}

\begin{figure}
\begin{subfigure}{0.4\textwidth}
\includegraphics[height=180px]{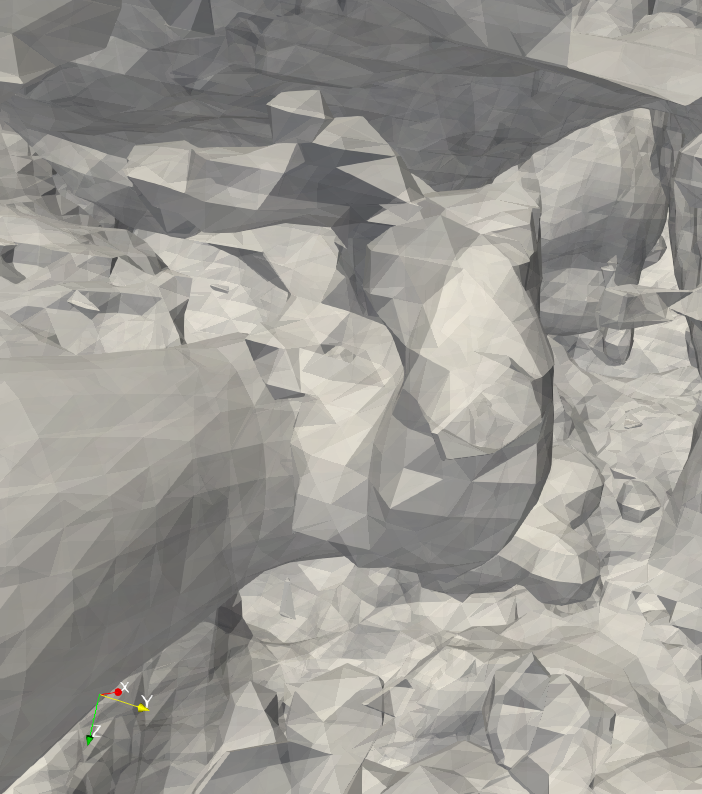}
\caption{}
\end{subfigure}
\quad
\begin{subfigure}{0.4\textwidth}
\includegraphics[height=180px]{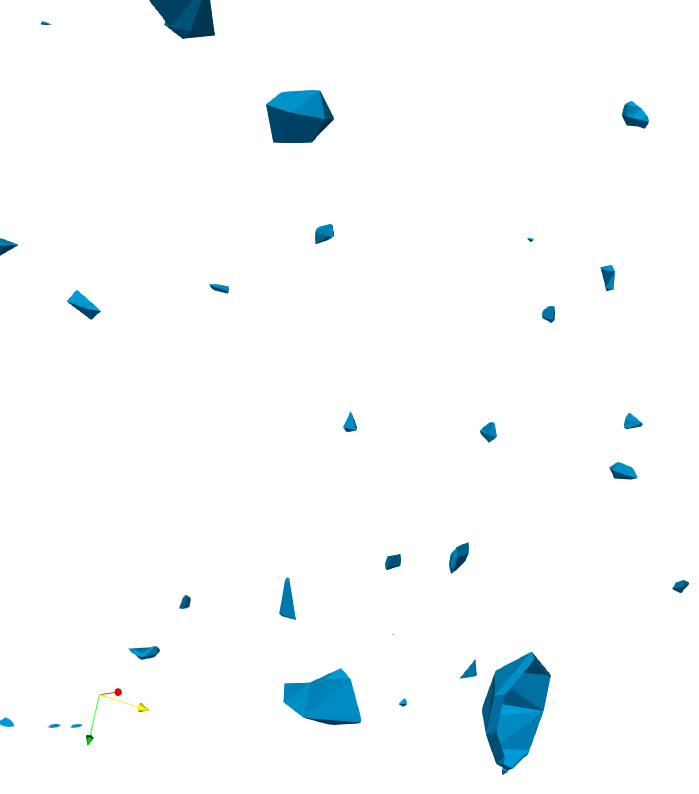}
\caption{}
\end{subfigure}
\quad
\begin{subfigure}{0.4\textwidth}
\includegraphics[height=179px]{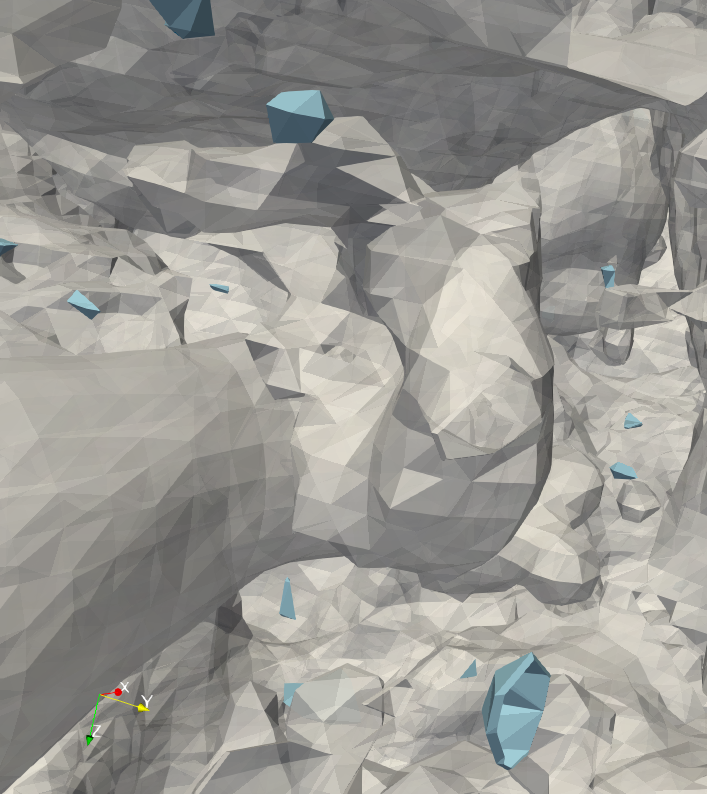}
\caption{}
\end{subfigure}
\caption{All four images are a close-up of the isosurface generated from visMale after slicing it across a plane. Created with isovalue 70.5 and filtered with a minimum size threshold of 5. (a) Isosurface after filtering out small components with scalar value above the isovalue. (b) Components that were removed when the filter removing small components below the isovalue was applied. (c) Images (a) and (b) combined.}
\label{fig:skullfilters}
\end{figure}

\begin{table*}
\centering
\begin{threeparttable}
\centering
\begin{minipage}{\textwidth}
\begin {tabular}{lrrrrr}%
\toprule Dataset & Isovalue & \# Total Cubes & \# Active Cubes & Marching Cubes Runtime & Filtering Alg. Runtime \\ \toprule %
\midrule abdominal\_stent&1350.5&45.1M&488K&0.423&1.36\\%
\midrule aneurysm&30.5&16.6M&163K&0.14&0.374\\%
\midrule bonsai&50.5&16.6M&305K&0.171&0.405\\%
\midrule carp&600.5\tnote {\fnII }&33.2M&450K&0.329&0.813\\%
\midrule carp&1150.5\tnote {\fnIII }&33.2M&662K&0.312&0.859\\%
\midrule colon\_prone&1500.5&120M&1433K&1.156&4.062\\%
\midrule colon\_supine&1500.5&111M&1392K&1.031&3.484\\%
\midrule lobster&20.5&5.33M&239K&0.078&0.141\\%
\midrule MRIwoman&1100.5&7.02M&599K&0.125&0.171\\%
\midrule skull&40.5&16.6M&959K&0.312&0.407\\%
\midrule visMale&55.5\tnote {\fnII }&8.26M&279K&0.11&0.187\\%
\midrule visMale&70.5\tnote {\fnIII }&8.26M&453K&0.125&0.22\\\bottomrule %
\end {tabular}%

  \begin{tablenotes}
\item [\fnII] Isovalue for skin.
\item [\fnIII] Isovalue for skeleton.
\end{tablenotes}
\end{minipage}
\caption{Running time in seconds to finish the Marching Cubes algorithm and the Filtering Algorithm on various datasets. Performed with minimum size threshold of 5. Each entry in the Filtering Algorithm Runtime column represents the time it takes to finish both the union-find on the scalar grid and the filtering of the small components.}
\labeltable{runtime}
\end{threeparttable}
\end{table*}

All the analyzed datasets measured the density,
in some form, of some object.
Thus, the interior of the object/surface is represented 
by high isovalues while the exterior is represented 
by very low (zero or near zero) isovalues.

Because object interiors are represented by high isovalues,
small connected isosurface components that are visible in the visualization
are removed by filtering small connected components
of scalar values GREATER than the isovalue.
On the other hand, small connected isosurface components 
that lie in ther interior of some larger component,
are only removed by filtering small connected components
of scalar values LESS than the isovalue.
Figure~\ref{fig:skullfilters} shows small connected isosurface components 
inside the skull that are not removed
when filtering small components with scalar value ABOVE the isovalue.
Note that the isosurface in Figure~\ref{fig:skullfilters}
was cropped to show the small components
that lie ``inside'' the larger main component of the skull.
These small components are filtered
if small components with scalar value BELOW the isovalue
are removed.

\section{Timing Analysis}
\labelsec{timing analysis}

All datasets are filtered with minimum size threshold 5. Refer to Table \reftable{runtime} below. From Algorithm \refalg{merge_components} and Section 3, we see that the time complexity of both algorithms are $O(T)$, where $T$ is the total number of cubes in the grid. Indeed, it can be seen that the runtimes of both algorithms increase almost linearly as the number of total cubes increases.

\section{Conclusion}
\labelsec{conclusion}

While this algorithm does successfully filter out small components out of the isosurface, creating a much less noisy object, there are still a few drawbacks with this method. One issue is that we cannot always determine whether a small particle is truly noise or part of an object whose isosurface is separated into many small pieces. Hence we may be removing components that are part of the isosurface itself, reducing the faithfulness of the filtered object to the original.

\bibliographystyle{abbrv-doi}

\bibliography{IsoComponentFiltering}

\end{document}